\documentclass[aps,prd,reprint,amsmath,amssymb,superscriptaddress,floatfix,twocolumn]{revtex4}
\usepackage{graphicx}
\usepackage{amsfonts}
\usepackage{comment}
\usepackage{amsmath}
\usepackage{subfigure}
\usepackage{tikz,mathpazo}
\usepackage{amssymb}
\usepackage{rotating}
\usepackage{booktabs}
\usepackage{xcolor}
\usepackage{soul}
\usepackage{color}
\usepackage{slashed}
\usepackage{multirow}
\usepackage{makecell}
\usepackage{epsf}
\usepackage{ulem}
\usepackage{cancel}
\usepackage{color,bm}
\usepackage{bm}
\usepackage{mathtools}

\usepackage[colorinlistoftodos]{todonotes}

\usepackage[colorlinks=true,citecolor=cyan,urlcolor=blue,bookmarks=true,bookmarks=true,bookmarksopen=true,bookmarksnumbered=true,bookmarksopenlevel=3]{hyperref}

\usepackage[compat=1.1.0]{tikz-feynman}

\definecolor{airforceblue}{rgb}{0.36, 0.54, 0.66}
\definecolor{steelblue}{rgb}{0.27, 0.51, 0.71}
\definecolor{amber}{rgb}{1.0, 0.49, 0.0}

\allowdisplaybreaks[4]

\begin{document}

\title{Theoretical study in the $\bar{B}^0 \to J/\psi \bar{K}^{*0} K^0$ and $\bar{B}^0 \to J/\psi f_1(1285)$ decays}
\author{DaZhuang He}
\author{Xuan Luo}
\author{YiLing Xie}
\author{{Hao Sun}\footnote{Corresponding author: haosun@dlut.edu.cn}}
\affiliation{Institute of Theoretical Physics, School of Physics, Dalian University of Technology, \\ No.2 Linggong Road, Dalian, Liaoning, 116024, P.R.China}
\date{\today}

\begin{abstract}  
\vspace{0.5cm}
We study the decay processes of $\bar{B}^0 \to J/\psi \bar{K}^{*0} K^0$ and $\bar{B}^0 \to J/\psi f_1(1285)$ to analyse the $f_1(1285)$ resonance.
By the calculation within chiral unitary approach where $f_1(1285)$ resonance is dynamically generated from the $K^*\bar{K}-c.c.$ interaction, 
we find that the $\bar{K}^{*0} K^0$ invariant mass distribution has a clear broad peak. 
Such broad peak has been understood as the signal of the $f_1(1285)$.
Finally, we obtain a theoretical result $R_t=\Gamma_{\bar{B}^0 \to J/\psi \bar{K}^{*0} K^0}/\Gamma_{\bar{B}^0 \to J/\psi f_1(1285)}$ 
which is expected to be compared with the experimental data.
\end{abstract}
\maketitle
\setcounter{footnote}{0}

\section{Introduction}
\label{I}

One of the main goals of the hadron physics is to unravel the nature of mesons or baryons. 
With the discovery of more and more new particles, the traditional quark model is not enough for us to understand the nature of particles. 
The general examples are $a_0(980)$ and $f_0(980)$, and several perspectives were presented about the nature of these: 
$q\bar{q}$ states \cite{Li:2000dy,Godfrey:1998pd}, glueballs, multiquark states or meson-meson molecules \cite{Klempt:2007cp}.

In very low energy regions, the strong interactions of the light pseudoscalar particles can be simplified by considering global symmetry.
Thus, based on the chiral symmetry  of Quantum Chromodynamics (QCD) and the concept of effective field theory(EFT), 
a powerful theoretical tool Chiral Perturbation Theory ($\chi PT$) has been proposed \cite{Pich:1995bw,Bernard:1995dp}. 
The $\chi PT$ can be used for the study of QCD at low energies($\leq 500$ MeV) by means of chiral Lagrangians using fields related to mesons and baryons. 
Further, the application of the unitary approaches in $\chi PT$, unitary chiral approach($U\chi PT$), provides a possibility to study strong interactions 
at higher energy($\ge 500$ MeV) by $\chi PT$. The $U\chi PT$ has been successfully applied to study meson-meson and meson-baryon interactions,
and also gives a new perspective to understand the nature of particles, since several resonances can be understood as dynamically generated. 
Such as, the $a_0(980)$ and $f_1(1285)$ have been viewed as dynamically generated from the interaction of $\bar{K} K$, 
$\pi \eta$ \cite{Oller:1997ti,Locher:1997gr,Xie:2016evi,Molina:2019udw,Duan:2020vye,Wang:2020pem} and $K^*\bar{K}-c.c.$ \cite{Xie:2015lta,Miyahara:2015cja} respectively. 
As for the baryon case, the $\Lambda(1405)$ can be shown as dynamically generated from the interaction of 
$\bar{K} N$ and $\eta \Lambda$ \cite{Oller:2000ma,Xie:2017gwc,Roca:2015tea,Miyahara:2015cja}, $\cdots$. 
In this paper, we focus on the $f_1(1285)$ as dynamically generated.

The $f_1 (1285)$ resonance is an axial-vector state with quantum numbers $I^G (J^{P C}) = 0^+ (1^{++})$, 
mass $M_{f_1} = 1281.9\pm 0.5$ MeV and total decay width $\Gamma_{f_1} = 22.7 \pm 1.1$ MeV \citep{Zyla:2020zbs}. 
It's hard for $f_1 (1285)$ to decay to two-body, since the mass of the $f_1(1285)$ is $100$ MeV below the $K^*\bar{K}$ threshold. 
The main decay modes of $f_1(1285)$ are the $\eta \pi \pi$ ($52.2\%$) and $4 \pi$ ($32.7\%$). 
In Ref.\citep{Roca:2005nm}, the authors found a pole of the amplitude of pseudoscalar-vector meson interaction 
in the complex plane at the $1288-i0$ MeV and assigned it to the $f_1(1285)$ resonance. 
At the same time, similar results were obtained in \cite{Zhou:2014ila} when including higher order Lagrangians.
Thus, in the chiral unitary approach, the $f_1(1285)$ can be qualified as dynamically generated from the pseudoscalar-vector meson interaction.
Although, the PDG \cite{Zyla:2020zbs} has reported "not seen" for the $K^*\bar{K}$ decay mode of the $f_1(1285)$, inspired by the $K \bar{K} \pi$ decay mode of the $f_1(1285)$, 
the authors in Ref.\cite{Aceti:2015pma} explained the possibility of the $K^*\bar{K}$ decay mode of the $f_1(1285)$ 
and gave the coupling $g_{f_1}=7555$ MeV of the $f_1(1285)$ to $K^*\bar{K}$ channel. 
Some works in Refs. \cite{Debastiani:2016xgg,Baguena:2017sju} also  show the possibility of the $K^*\bar{K}$ decay mode of the $f_1(1285)$.

In Refs.\citep{Xie:2015lta, Molina:2016pbg}, the role of the $f_1(1285)$ that was dynamically generated from the interaction of $K^*\bar{K}-c.c.$,
has been studied in the $J/\psi \to \phi f_1(1285)$ and $\bar{B}_s^0 \to J/\psi f_1(1285)$ decay processes, respectively. 
Unlike the latter which showed only one tail of the resonance peak, the former showed a clear threshold enhancement of invariant mass distribution of $\bar{K}K^*$, 
which is caused by the production of the $f_1(1285)$. In Ref.\cite{Aceti:2015zva}, they obtained a consistent result with the experiment 
for the $f_1(1285) \to \pi^0 a_0(980)$ decay branching fraction, where the $f_1(1285)$, $a_0(980)$ are also treated as dynamically generated. 
Also, the process $\tau \to f_1(1285)\pi \nu_\tau$ has been researched by considering a triangle loop mechanism with $K^*(\bar{K}^*)$ and $\bar{K}(K)$ as internal lines \cite{Oset:2018zgc}. 

On the other hand, in the experiments, the $B^0_{(s)}$ provides a good platform for the study of 
some resonances $f_0, a_0, f_1, \cdots$ \cite{Bayar:2014qha,Xie:2014gla,Liang:2014ama,Dai:2015bcc}. 
Following the original ideas in Refs.\citep{Xie:2015lta,Molina:2016pbg}, we study the decays of $\bar{B}^0\to J/\psi \bar{K}^{*0} K^0$ 
and $\bar{B}^0\to J/\psi f_1(1285)$ decay by considering that the $f_1(1285)$ resonance is dynamically generated from $K^* \bar{K}$ interaction.
Notice although the channel $\bar{B}^0\to J/\psi \bar{K}^{*0} K^0$ cannot be observed directly, it can be related to  $\bar{B}^0 \to J/\psi K^0 K^- \pi^+ +c.c.$ 
by decay of the $K^* \to K \pi$, which was measured by LHCb Collaboration with $Br(B^0 \to J/\psi K^0 K^- \pi^+ +c.c.)= 2.1 \times 10^{-5}$ \cite{Aaij:2014naa}.
The branching ratio of $Br(B^0 \to J/\psi f_1(1285))= (8.4 \pm 2.1)\times 10^{-6}$ was also obtained 
from the measurement of $[\Gamma(B^0\to J/\psi f_1(1285))/\Gamma_{total}]\times [B(f_1(1285) \to 2\pi^+ 2\pi^-)]$ by the LHCb Collaboration \cite{Lees:2012cr,Aaij:2013rja}. 
We can therefore define the physical quantity $R_\Gamma=\frac{d\Gamma_{\bar{B}^0 \to J/\psi \bar{K}^{*0} K^0}/dM_{inv}(\bar{K}^{*0} K^0)}{\Gamma_{\bar{B}^0 \to J/\psi f_1(1285)}}$ 
without any other free parameter dependence except the energy scale, and show the distribution of $\bar{K}^{*0} K^0$ invariant mass. 
Moreover, we can obtain a theoretical result $R_t=\Gamma_{B^0 \to J/\psi \bar{K}^{*0} K^0}/\Gamma_{\bar{B}^0 \to J/\psi f_1(1285)}$ 
which is possible to be compared with the experimental data at a certain energy scale. 

This paper is organized as follows. 
In Sec.\ref{II}, we give the theoretical formalism for the weak decay of $b$ quark and the  hadronization of quark pairs. 
In Sec.\ref{III}, the Lagrangian of the processes $VP \to VP$ and the detailed calculations are presented. 
In Sec.\ref{IV}, the numerical results and discussions are presented. 
Finally, a short summary is given in Sec.\ref{V}.

\section{Formalism}
\label{II}

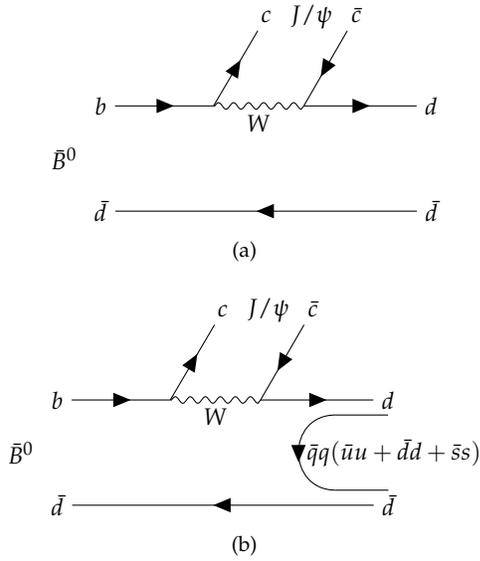
\begin{figure}[h]
\begin{center} 
	\subfigure[]{
		\begin{tikzpicture}
		\begin{feynman}
		\vertex(a1){\( b \)};
		\vertex[below=0.7cm of a1] (f1);
		\vertex[left=0.2cm of f1] (f2){\( \bar{B}^0 \)};
	
		\vertex[right=1.5cm of a1] (a2);
		\vertex[right=0.7cm of a2] (a21);
		\vertex[above=1cm of a21] (a22){\(c \)};
		\vertex[right=0.6cm of a22] (l2){\(J/\psi \)};
		\vertex[right=1.2cm of a2] (a3);
		\vertex[right=0.7cm of a3] (a31);
		\vertex[above=1cm of a31] (a32){\(\bar{c} \)};
		\vertex[right=1.5cm of a3] (a4){\(d \)};
	
		\vertex[below=1.4cm of a1] (b1){\( \bar{d} \)};
		\vertex[right=4.4cm of b1] (b2){\( \bar{d} \)};
		\vertex[right=2.2cm of b1] (l1);
	
		\vertex[below=0.2cm of a4] (c1);
		\vertex[left=0.2cm of c1] (c11);
		\vertex[above=0.2cm of b2] (c2);
		\vertex[left=0.2cm of c2] (c21);
		\diagram*[baseline=(a5.base)] {
		{
	 	(a1) --[fermion] (a2) --[photon,edge label'=\( W \)] (a3)--[fermion](a4),
		},
		(b2)--[fermion](b1),
		(a2)--[fermion](a22),
		(a32)--[fermion](a3),
		};
		\end{feynman}
		\end{tikzpicture}
	} 
	\subfigure[]{
	\begin{tikzpicture}
	\begin{feynman}
	\vertex(a1){\( b \)};
	\vertex[below=0.7cm of a1] (f1);
	\vertex[left=0.2cm of f1] (f2){\( \bar{B}^0 \)};
	
	\vertex[right=1.5cm of a1] (a2);
	\vertex[right=0.7cm of a2] (a21);
	\vertex[above=1cm of a21] (a22){\(c \)};
	\vertex[right=0.6cm of a22] (l2){\(J/\psi \)};
	\vertex[right=1.2cm of a2] (a3);
	\vertex[right=0.7cm of a3] (a31);
	\vertex[above=1cm of a31] (a32){\(\bar{c} \)};
	\vertex[right=1.5cm of a3] (a4){\(d \)};
	
	\vertex[below=1.4cm of a1] (b1){\( \bar{d} \)};
	\vertex[right=4.4cm of b1] (b2){\( \bar{d} \)};
	\vertex[right=2.2cm of b1] (l1);

	\vertex[below=0.2cm of a4] (c1);
	\vertex[left=0.7cm of c1] (c11);
	\vertex[above=0.2cm of b2] (c2);
	\vertex[left=0.7cm of c2] (c21);
	\diagram*[baseline=(a5.base)] {
	{
	 (a1) --[fermion] (a2) --[photon,edge label'=\( W \)] (a3)--[fermion](a4),
	},
	(b2)--[fermion](b1),
	(a2)--[fermion](a22),
	(a32)--[fermion](a3),
	(c1)--(c11),
	(c2)--(c21),
	(c11)--[fermion, out=180, in=180, looseness=1.68,edge label=\(\bar{q}q(\bar{u}u+\bar{d}d+\bar{s}s)\)](c21),
	};
	\end{feynman}
	\end{tikzpicture} 
	}
	\caption{The $\bar{B}^0 \to J/\psi f_1(1285)$ decay process at the quark level. 
	(a) Elementary quark  arrangement for the weak decay of $b$ quark. 
	(b) Hadronization of the $d\bar{d}$ component.}
	\label{fig1}
\end{center}
\end{figure}

In this section we draw the pictures of the weak decay of $b$ quark and the  hadronization. 
The reaction process at the quark level as shown in the FIG.\ref{fig1} can be divided into three steps. 
The first part is that the $b$ quark of the $\bar{B}^0$ converts to a $c\bar{c}$ and a $d$ quark via weak interaction, 
where two vertex $bcW$ and $dcW$ involved by the weak decay are Cabibbo suppressed \cite{Chau:1987tk,Liang:2014tia} as shown in the FIG.\ref{fig1}(a). 
In the next step, the $c\bar{c}$ pair from decay of the $b$ quark forms a $J/\psi$ 
and the $d\bar{d}$ hadronizes a pair of vector-pseudoscalar mesons as depicted in FIG.\ref{fig1}(b). 
Following Refs.\cite{Liang:2017ijf,Molina:2016pbg,Miyahara:2015cja}, for the $d\bar{d}$ hadronization, 
we need to introduce a $\bar{q}q$ pair with the quantum numbers of the vacuum, $\bar{u}u+\bar{d}d+\bar{s}s$, 
and then two pairs of $q\bar{q}$(vector-pseudoscalar mesons) arise. The hadronization process can be written 
\begin{equation}\label{eq1}
\begin{aligned}
d(\bar{u}u+\bar{d}d+\bar{s}s)\bar{d}=\sum_{i=1}^3 M_{2i}M_{i2}
\end{aligned}
\end{equation}
where $i$ denotes the quarks $u,d,s$, and $M$ is the $q\bar{q}$ matrix in SU(3) group
\begin{equation}\label{eq2}
M=q\bar{q}=
	\begin{pmatrix}
	u\\
	d\\
	s	
	\end{pmatrix}
	\begin{pmatrix}
	\bar{u} & \bar{d} & \bar{s} 
	\end{pmatrix}
	=\begin{pmatrix}
	u\bar{u}& u\bar{d}& u\bar{s}\\
	d\bar{u}& d\bar{d}& d\bar{s}\\
	s\bar{u}& s\bar{d}& s\bar{s} 
\end{pmatrix}_.
\end{equation}
The matrix $M$ can also be rewritten in terms of pseudoscalar meson matrix $M_P$ or vector meson matrix $M_V$, 
\begin{equation}\label{eq3}
\begin{aligned}
&M_P=\begin{pmatrix}
\frac{\pi^0}{\sqrt{2}}+\frac{\eta_8}{\sqrt{6}} &\pi^+ &K^+\\
 \pi^-&-\frac{\pi^0}{\sqrt{2}}+\frac{\eta_8}{\sqrt{6}} &K^0\\
 K^-& \bar{K}^0& -\frac{2}{\sqrt{6}}\eta_8
\end{pmatrix}_, \qquad\\
&M_V=\begin{pmatrix}
	\frac{\rho^0}{\sqrt{2}}+\frac{\omega}{\sqrt{2}}& \rho^+& K^{*+}\\
	\rho^-&-\frac{\rho^0}{\sqrt{2}}+\frac{\omega}{\sqrt{2}} & K^{*0}\\
	K^{*-}& \bar{K}^{*0}& \phi 
	\end{pmatrix}_.
\end{aligned}
\end{equation}
Eq.(\ref{eq1}) can be rewritten with regard to $M_V M_P $ or $M_P M_V$ 
\begin{equation}\label{eq4}
\begin{aligned}
 M_V M_P \pm M_P M_V ,
\end{aligned}
\end{equation} 
where the $\pm$ corresponds to $\mathcal{C}=\mp 1$($\mathcal{C}$ parity), respectively \cite{Molina:2016pbg}. Now, the hadronization process of $d\bar{d}$ gives
\begin{equation}\label{eq5}
\begin{aligned}
d(u\bar{u}&+d\bar{d}+s\bar{s}) \bar{d}=( M_V M_P - M_P M_V )_{22}\\
&= K^{*0} \bar{K}^0-\bar{K}^{*0}K^0  +\rho^- \pi^+  -\rho^+ \pi^-  \\
&+\left(\frac{\pi^0}{\sqrt{2}}+\frac{\eta_8}{\sqrt{6}} \right)\left(-\frac{\rho^0}{\sqrt{2}}+\frac{\omega}{\sqrt{2}} \right)\\
&-\left(-\frac{\rho^0}{\sqrt{2}}+\frac{\omega}{\sqrt{2}} \right)\left(\frac{\pi^0}{\sqrt{2}}+\frac{\eta_8}{\sqrt{6}} \right) .
\end{aligned}
\end{equation}

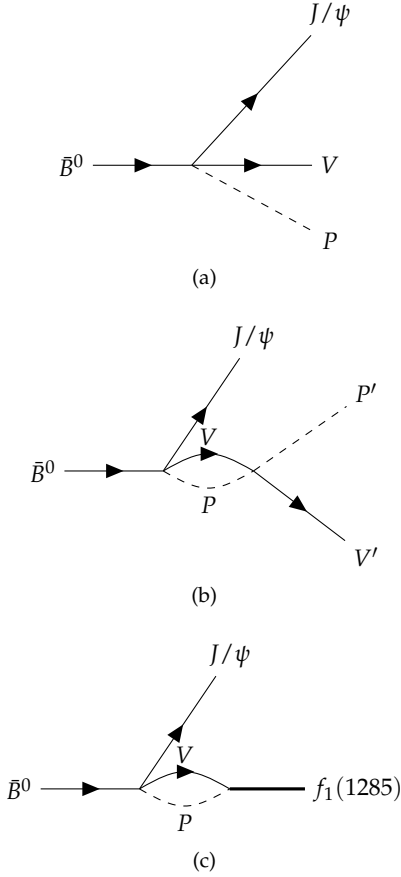
\begin{figure}[h]
\begin{center} 
	\subfigure[ ]{
	\begin{tikzpicture}
	\begin{feynman}
	\vertex(a1){\( \bar{B}^0 \)};
	\vertex[right=1.6cm of a1] (a2);
	\vertex[right=1.6cm of a2] (a3){\( V \)};
	\vertex[above=2cm of a3] (a4){\( J/\psi \)};
	\vertex[below=1 cm of a3] (a5){\( P \)};
	\diagram*[baseline=(a3.base)] {
	{
	 (a1) --[fermion] (a2) --[fermion] (a3),
	},
	(a2)--[fermion](a4),
	(a2)--[scalar](a5),
	};
	\end{feynman}
	\end{tikzpicture}
	} 
	\subfigure[ ]{
	\begin{tikzpicture}
	\begin{feynman}
	\vertex(a1){\( \bar{B}^0 \)};
	\vertex[right=1.6cm of a1] (a2);
	\vertex[right=1.2cm of a2] (a3);
	\vertex[above=1.5cm of a3] (a4){\( J/\psi \)};
	\vertex[right=1.5 cm of a3] (a5);
	\vertex[above=0.8 cm of a5] (a6){\( P'\)};
	\vertex[below=0.9 cm of a5] (a7){\( V'\)};
	\diagram*[baseline=(a3.base)] {
	{
	 (a1) --[fermion] (a2),
	},
	(a2)--[fermion, out=30, in=150, looseness=1.3,edge label=\( V \)](a3),
	(a2)--[scalar, out=330, in=210, looseness=1.3,edge label'=\(P\)](a3),
	(a2)--[fermion](a4),
	(a3)--[scalar](a6),
	(a3)--[fermion](a7),
	};
	\end{feynman}
	\end{tikzpicture} }
	\subfigure[ ]{
	\begin{tikzpicture}
	\begin{feynman}
	\vertex(a1){\( \bar{B}^0 \)};
	\vertex[right=1.6cm of a1] (a2);
	\vertex[right=1.2cm of a2] (a3);
	\vertex[above=1.5cm of a3] (a4){\( J/\psi \)};
	\vertex[right=1 cm of a3] (a5){\( f_1(1285) \)};
	\vertex[below=1.2cm of a5] (a6);
	\diagram*[baseline=(a3.base)] {
	{
	 (a1) --[fermion] (a2),
	},
	(a2)--[fermion, out=30, in=150, looseness=1.3,edge label=\( V \)](a3),
	(a2)--[scalar, out=330, in=210, looseness=1.3,edge label'=\(P\)](a3),
	(a2)--[fermion](a4),
	(a3)--[very thick](a5),
	};
	\end{feynman}
	\end{tikzpicture} }
	\caption{Diagrammatic representation for the $\bar{B}^0 \to J/\psi VP$ process. 
	(a) The reaction at the tree level. 
	(b) The rescattering between pseudoscalar mesons and vector mesons. 
	(c) Production process of the $\bar{B}^0 \to J/\psi f_1(1285)$ decay.}
	\label{fig2}
\end{center}
\end{figure}

In the last step from FIG.\ref{fig2}(b) which will be explained in the following section,
we need pseudoscalar $P$ and vector mesons $V$ for rescattering to form the molecular resonance $f_1(1285)$ 
with quantum numbers $I^G(J^{PC})=0^+(1^{++})$ \cite{Zyla:2020zbs} since $\mathcal{C} V=-\bar{V}$ and $\mathcal{C} P=\bar{P}$ for $M_V$ and $M_P$, respectively. 
Thus, in order to generate the resonance $f_1(1285)$ we must take the C-parity positive combination between the $V$ and $P$, 
and the Eq.(\ref{eq5}):
\begin{equation}
\begin{aligned}\label{eq6}
|MM \rangle=|K^{*0} \bar{K}^0-\bar{K}^{*0}K^0  +\rho^- \pi^+  -\rho^+ \pi^-  \rangle.
\end{aligned}
\end{equation}

\section{The processes $\bar{B}^0 \to J/\psi \bar{K}^{*0} K^0$ and $\bar{B}^0 \to J/\psi f_1(1285)$}
\label{III}

In this section, we will calculate the invariant mass distribution of $\bar{K}^{*0} K^0$ in  decay process $\bar{B}^0 \to J/\psi \bar{K}^{*0} K^0$ and the decay width of $\bar{B}^0 \to J/\psi f_1(1285)$.
According to the chiral unitary approach \cite{Roca:2005nm}, the molecular resonance $f_1(1285)$ can be regarded as dynamically generated from the interaction of $K^* \bar{K}-c.c.$. 
Meanwhile, the process of $VP \to VP$ rescattering is shown in  FIG.\ref{fig2}(b). 
In the local hidden gauge approach \cite{Bando:1984ej,Meissner:1987ge,Birse:1996hd,Bando:1987br}, 
the Lagrangian for the interaction of two pseudoscalar and two vector mesons is  
\begin{eqnarray}\label{eq7}
\mathcal{L}_{VVPP}=-\frac{1}{4f_\pi ^2} Tr \left( \left[V^\mu,\partial^\nu V_\mu  \right] \left[P,\partial_\nu P  \right]  \right)
\end{eqnarray}    
where $P$ and $V$ are given in Eq.(\ref{eq3}), $f_\pi$ is the decay constant of pion with $f_\pi=92.4$ MeV, 
$Tr$ represents the $SU(3)$ trace. The Lagrangian of Eq.(\ref{eq7}) will lead to the relevant the  $s$-wave projection of the scattering amplitude, or to say the  $V$-matrix \citep{Roca:2005nm}
\begin{eqnarray}\label{eq8} \nonumber
V_{ij}(s)&=&-\frac{\epsilon \cdot \epsilon'}{8f_\pi^2} C_{ij} \bigg[3s-(M^2+m^2+M'^2+m'^2)\\
			&-&\frac{1}{s}(M^2-m^2)(M'^2-m'^2) \bigg]
\end{eqnarray}
where $\epsilon (\epsilon')$ represents the polarization vector of the incoming(outgoing) vector meson. 
For convenience, $\epsilon \cdot \epsilon'$ is calculated as a prefactor below. 
The $M(M')$, $m(m')$ stand for the masses of initial(final) vector mesons and initial(final) pseudoscalar mesons, respectively. 
The indices $i$ and $j$ correspond to the initial and final $V P$ states, respectively. 
There are different $C_{ij}$ coefficients for different isospin basis $(S,I)$, the results of $C_{ij}^{\footnotemark[1]}$ have been listed in Ref.\citep{Roca:2005nm}.

The rescattering amplitude $T$ for the decay of $\bar{B}^0 \to J/\psi \bar{K}^{*0} K^0 $ as shown in the FIG.\ref{fig2}(b) 
was obtained by solving the Bethe-Salpeter equation in coupled channels \cite{Roca:2005nm}
\begin{eqnarray}\label{eq9}
T=[1+V\hat{G}]^{-1}(-V)\vec{\epsilon}\cdot \vec{\epsilon}^\prime
\end{eqnarray}
where $\hat{G}=G\left(1+\frac{1}{3}\frac{q^2}{M^2} \right)$, and $G$ is the two meson propagator loop function including a vector and a pseudoscalar:
\begin{eqnarray}\label{eq10}
G(\sqrt{s})=i\int \frac{d^4q}{(2\pi)^4} \frac{1}{(P-q)^2-M^2+i\epsilon}\frac{1}{q^2-m^2+i\epsilon}, ~
\end{eqnarray} 
where $m$ and $M$ are the masses of the pseudoscalar and vector mesons, respectively. 
Some amplitudes and $V$-matrices have been listed in Ref.\cite{Wang:2020pem, Toledo:2020zxj}.

The divergence of loop function Eq.(\ref{eq10}) can be regularized by means of the dimensional regularization \cite{Geng:2015yta,Roca:2005nm} 
or cut-off scheme \cite{Aceti:2015zva}, one has in the former scheme
\begin{eqnarray}\label{eq11} \nonumber
G^D(\sqrt{s})&=&\frac{1}{16\pi^2} \bigg\{ a(\mu)+\ln{\frac{M^2}{\mu^2}}+\frac{m^2-M^2+s}{2s}\ln{\frac{m^2}{M^2}}\\ \nonumber
	&+&\frac{|\vec{q}|}{\sqrt{s}}[\ln{(s-(M^2-m^2)+2|\vec{q}|\sqrt{s})}\\ \nonumber
	&+&\ln{(s+(M^2-m^2)+2|\vec{q}|\sqrt{s})}\\ \nonumber
	&-&\ln{(-s+(M^2-m^2)+2|\vec{q}|\sqrt{s})}\\
	&-&\ln{(-s-(M^2-m^2)+2|\vec{q}|\sqrt{s})}] \bigg\}.
\end{eqnarray}
Also, in the cut-off scheme we have
\begin{equation}\label{eq12}
\begin{aligned}
G(\sqrt{s})=&\int_0^{q_{max}} \frac{d^3 \vec{q}}{(2\pi)^3} \frac{\omega_1+\omega_2}{2\omega_1\omega_2}\frac{1}{s-(\omega_1+\omega_2)^2+i\epsilon}\\
\omega_1=&\sqrt{|\vec{q}|^2+m^2} \qquad \omega_2=\sqrt{|\vec{q}|^2+M^2}_.
\end{aligned}
\end{equation} 
where the  $\mu$ and $a(\mu)$ are the  renormalization  scale and subtraction constant, respectively.
And, the $\vec{q}$ stands for the three-momentum
of the vector or pseudoscalar meson in the center of mass frame and is given by

\begin{equation}\label{eq13}
|\vec{q}|=\frac{1}{2\sqrt{s}}\sqrt{[s-(M+m)^2][s-(M-m^2)]}.
\end{equation} 
The $s$ is the invariant mass of vector and pseudoscalar mesons in the loop of FIG.\ref{fig2}(b).

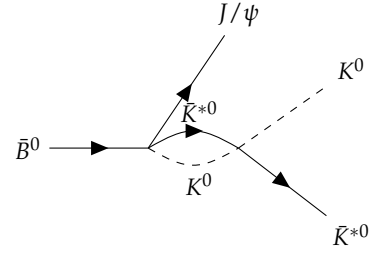
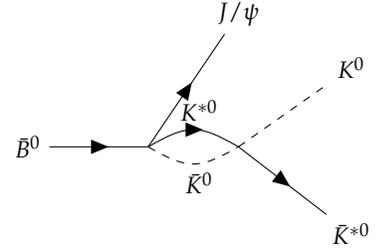
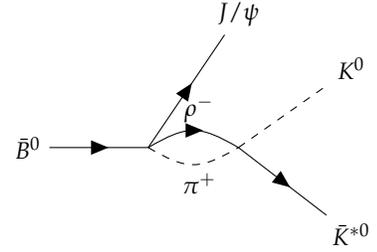
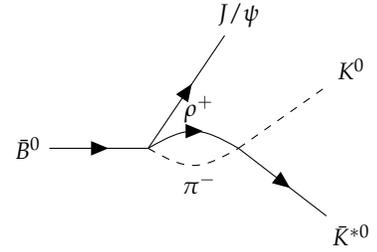
\begin{figure}[h]
\begin{center}
	\subfigure[ ]{
	\begin{tikzpicture}
	\begin{feynman}
	\vertex(a1){\( \bar{B}^0 \)};
	\vertex[right=1.6cm of a1] (a2);
	\vertex[right=1.2cm of a2] (a3);
	\vertex[above=1.5cm of a3] (a4){\( J/\psi \)};
	\vertex[right=1.5 cm of a3] (a5);
	\vertex[above=0.8 cm of a5] (a6){\( K^0 \)};
	\vertex[below=0.9 cm of a5] (a7){\( \bar{K}^{*0} \)};
	\diagram*[baseline=(a3.base)] {
	{
	 (a1) --[fermion] (a2),
	},
	(a2)--[fermion, out=30, in=150, looseness=1.3,edge label=\( \ \   \bar{K}^{*0} \)](a3),
	(a2)--[scalar, out=330, in=210, looseness=1.3,edge label'=\(\ \ K^0\)](a3),
	(a2)--[fermion](a4),
	(a3)--[scalar](a6),
	(a3)--[fermion](a7),
	};
	\end{feynman}
	\end{tikzpicture} } 
	\subfigure[ ]{
	\begin{tikzpicture}
	\begin{feynman}
	\vertex(a1){\( \bar{B}^0 \)};
	\vertex[right=1.6cm of a1] (a2);
	\vertex[right=1.2cm of a2] (a3);
	\vertex[above=1.5cm of a3] (a4){\( J/\psi \)};
	\vertex[right=1.5 cm of a3] (a5);
	\vertex[above=0.8 cm of a5] (a6){\( K^0 \)};
	\vertex[below=0.9 cm of a5] (a7){\( \bar{K}^{*0} \)};
	\diagram*[baseline=(a3.base)] {
	{
	 (a1) --[fermion] (a2),
	},
	(a2)--[fermion, out=30, in=150, looseness=1.3,edge label=\( \ \ K^{*0}  \)](a3),
	(a2)--[scalar, out=330, in=210, looseness=1.3,edge label'=\(\ \ \bar{K}^0\)](a3),
	(a2)--[fermion](a4),
	(a3)--[scalar](a6),
	(a3)--[fermion](a7),
	};
	\end{feynman}
	\end{tikzpicture} }
   \subfigure[ ]{
	\begin{tikzpicture}
	\begin{feynman}
	\vertex(a1){\( \bar{B}^0 \)};
	\vertex[right=1.6cm of a1] (a2);
	\vertex[right=1.2cm of a2] (a3);
	\vertex[above=1.5cm of a3] (a4){\( J/\psi \)};
	\vertex[right=1.5 cm of a3] (a5);
	\vertex[above=0.8 cm of a5] (a6){\( K^0 \)};
	\vertex[below=0.9 cm of a5] (a7){\( \bar{K}^{*0} \)};
	\diagram*[baseline=(a3.base)] {
	{
	 (a1) --[fermion] (a2),
	},
	(a2)--[fermion, out=30, in=150, looseness=1.3,edge label=\(\ \ \rho^- \)](a3),
	(a2)--[scalar, out=330, in=210, looseness=1.3,edge label'=\(\ \ \pi^+\)](a3),
	(a2)--[fermion](a4),
	(a3)--[scalar](a6),
	(a3)--[fermion](a7),
	};
	\end{feynman}
	\end{tikzpicture} } 
	\subfigure[ ]{
	\begin{tikzpicture}
	\begin{feynman}
	\vertex(a1){\( \bar{B}^0 \)};
	\vertex[right=1.6cm of a1] (a2);
	\vertex[right=1.2cm of a2] (a3);
	\vertex[above=1.5cm of a3] (a4){\( J/\psi \)};
	\vertex[right=1.5 cm of a3] (a5);
	\vertex[above=0.8 cm of a5] (a6){\( K^0 \)};
	\vertex[below=0.9 cm of a5] (a7){\( \bar{K}^{*0} \)};
	\diagram*[baseline=(a3.base)] {
	{
	 (a1) --[fermion] (a2),
	},
	(a2)--[fermion, out=30, in=150, looseness=1.3,edge label=\(\ \ \rho^+ \)](a3),
	(a2)--[scalar, out=330, in=210, looseness=1.3,edge label'=\(\ \ \pi^-\)](a3),
	(a2)--[fermion](a4),
	(a3)--[scalar](a6),
	(a3)--[fermion](a7),
	};
	\end{feynman}
	\end{tikzpicture} }
	\caption{The total contributions for the $\bar{K}^{*0} K^0$ production. Subfigures (a) and (d) have a minus sign corresponding with subfigures (b) and (c).}
	\label{fig3}
\end{center}
\end{figure}

When the final state is the $\bar{K}^{*0} K^0$, the total amplitude contributions correspond to FIG.\ref{fig3}. 
Eq.(\ref{eq6}) causes that subfigures (b) and (c) have a minus sign comparing with (a) and (d). 
Meanwhile, we can write the amplitude $\mathcal{U}_j$ for the transition $\bar{B}^0 \to J/\psi \bar{K}^{*0} K^0$
\begin{equation}\label{eq14}
\begin{aligned}
\mathcal{U}_j =& V_P \left( h_j+\sum_i h_i \hat{G}_i(M_{inv})t_{ij}(M_{inv}) \right)\\
\end{aligned}
\end{equation}
where the $\hat{G}$ is the loop function in Eq.(\ref{eq10}), 
the scattering matrix $t_{ij}$ is described in Eq.(\ref{eq9}) and $M_{inv}$ is the invariant mass of the vector-meson and pseudoscalar-meson in the final state. 
The $V_P$ is an overall factor, which includes Cabibbo-Kobayashi-Maskawa(CKM) matrix elements and kinematic prefactors. 
And the $V_P$ is a unknown quantity and it could be cancelled below. 
Meanwhile, in the Refs. \cite{Kang:2013jaa,Xie:2016evi,Daub:2015xja,Xie:2017gwc,Wang:2015uea} the authors take $V_P$ as a constant. 
Also, considering the Eq.(\ref{eq8}), for $h_{i(j)}$ we have
\begin{equation}\label{eq15}
\begin{aligned}
h_{K^{*0} \bar{K}^0}&=h_{\rho^- \pi^+}=1,  \quad h_{\bar{K}^{*0}K^0}=h_{\rho^+ \pi^-}=-1, \\
h_{K^{*+}K^-}&= h_{K^{*-} K^+}=h_{\rho^0 \pi^0}= h_{\rho^0 \rho^0}=0.
\end{aligned}
\end{equation}
Take notice that the above amplitude only holds for an s-wave with every intermediate particle being on mass-shell.

The amplitude Eq.(\ref{eq14}) for the $\bar{B}^0 \to J/\psi \bar{K}^{*0} K^0$ can be written as
\begin{eqnarray}\label{eq16} \nonumber
&&\mathcal{U}=V_P \big[-1 \\  \nonumber
      &-&G_{\bar{K}^{*0} K^0}(M_{inv}(\bar{K}^{*0} K^0))t_{\bar{K}^{*0} K^0 \to \bar{K}^{*0} K^0}(M_{inv}(\bar{K}^{*0} K^0))\\ \nonumber
	&+&G_{K^{*0} \bar{K}^0}(M_{inv}(\bar{K}^{*0} K^0))t_{K^{*0} \bar{K}^0 \to \bar{K}^{*0} K^0}(M_{inv}(\bar{K}^{*0} K^0))\\ \nonumber
	&+&G_{\rho^- \pi^+}(M_{inv}(\bar{K}^{*0} K^0))t_{\rho^- \pi^+ \to \bar{K}^{*0} K^0}(M_{inv}(\bar{K}^{*0} K^0))\\ \nonumber
	&-&G_{\rho^+ \pi^-}(M_{inv}(\bar{K}^{*0} K^0))t_{\rho^+ \pi^- \to \bar{K}^{*0} K^0}(M_{inv}(\bar{K}^{*0} K^0)) \big]. \\
\end{eqnarray}
By means of the following isospin multiplets
\begin{equation}\label{eq17}
\begin{aligned}
\begin{pmatrix}
K^+\\
K^0
\end{pmatrix}_,\quad
\begin{pmatrix}
\bar{K}^0\\
-K^-
\end{pmatrix}_,\quad
\begin{pmatrix}
-\rho^+\\
\rho^0\\
\rho^-
\end{pmatrix}_, \quad
\begin{pmatrix}
-\pi^+\\
\pi^0\\
\pi^-
\end{pmatrix}_,
\end{aligned}
\end{equation}
and the CG-coefficients, we have amplitude relations
\begin{equation}\label{eq18}
\begin{aligned}
t_{\bar{K}^{*0} K^0 \to \bar{K}^{*0} K^0}=&+\frac{1}{2}t_{\bar{K}^{*} K \to \bar{K}^{*} K}^{I=0}   \\
t_{K^{*0} \bar{K}^0 \to \bar{K}^{*0} K^0}=&-\frac{1}{2}t_{K^{*} \bar{K} \to \bar{K}^{*} K}^{I=0}  \\
t_{\rho^-\pi^+ \to \bar{K}^{*0} K^0}=&-\frac{1}{\sqrt{6}}t_{\rho \pi \to \bar{K}^{*} K}^{I=0}   \\
t_{\rho^+\pi^- \to \bar{K}^{*0} K^0}=&-\frac{1}{\sqrt{6}}t_{\rho \pi \to \bar{K}^{*} K.}^{I=0} \\
\end{aligned}
\end{equation}  
And we can rewrite the Eq.(\ref{eq16}) as
\begin{eqnarray}\label{eq19} \nonumber
&&\mathcal{U}=-V_P \big[ 1 \\ \nonumber
&+&\frac{1}{2} G_{\bar{K}^* K}(M_{inv}(\bar{K}^{*0} K^0)) t_{\bar{K}^* K \to \bar{K}^* K}^{I=0}(M_{inv}(\bar{K}^{*0} K^0)) \\ 
&+&\frac{1}{2} G_{K^* \bar{K}}(M_{inv}(\bar{K}^{*0} K^0)) t_{K^* \bar{K} \to \bar{K}^* K}^{I=0}(M_{inv}(\bar{K}^{*0} K^0))  \big]. ~~~~~ 
\end{eqnarray}
If the final state is chosen as $J/\psi K^{*0} \bar{K}^0 $, the amplitude for $\mathcal{U}_{J/\psi K^{*0} \bar{K}^0}$ has a minus sign compared with Eq.(\ref{eq19}).

With the above results, now the invariant mass distribution $d\Gamma/dM_{inv}$ can be written \cite{Zyla:2020zbs}
\begin{eqnarray}\label{eq20} \nonumber
\frac{d\Gamma}{dM_{inv}} &=&	\sum C_s^2 \frac{1}{(2\pi)^3}\frac{M_{inv}}{8M_{\bar{B}^0}^3} \\
 &\times& \int_{M_{J/\psi,\bar{K}^{*0}}^{min}}^{M_{J/\psi,\bar{K}^{*0}}^{max}} |\mathcal{U}|^2 M_{J/\psi,\bar{K}^{*0}} dM_{J/\psi,\bar{K}^{*0}},
\end{eqnarray}
the $\sum C_s^2$ represents the total polarization structure, which can be written as
\begin{equation}\label{eq21}
\begin{aligned}
\sum C_s^2=3+ \frac{\vec{p}_{J/\psi}^2}{m_{J/\psi}^2}+\frac{\vec{p}_{\bar{K}^{*0}}^2}{m_{\bar{K}^{*0}}},\\
\end{aligned}
\end{equation}
 where we make the approximation\cite{Liang:2017ijf,Sakai:2017hpg,Oset:2009vf} that the three-momenta of the vector mesons in the loop are on-shell and small compared to their masses. And we have
\begin{equation}
\begin{aligned}
\vec{p}_{J/\psi}= & \frac{\lambda^{\frac{1}{2}}(M_{\bar{B}^0}^2,M_{J/\psi}^2,M_{inv}^2)}{2M_{\bar{B}^0}},\\
\vec{p}_{\bar{K}^{*0}}=&\frac{\lambda^{\frac{1}{2}}(M_{inv}^2,M_{k^0}^2,M_{\bar{k}^{*0}}^2)}{2M_{inv}},
\end{aligned}
\end{equation}
where $\lambda(x, y, z)=x^2+y^2+z^2-2xy-2yz-2xz$ is the K\"ahl\'en function. 

\begin{figure}[h]
\includegraphics[scale=0.6]{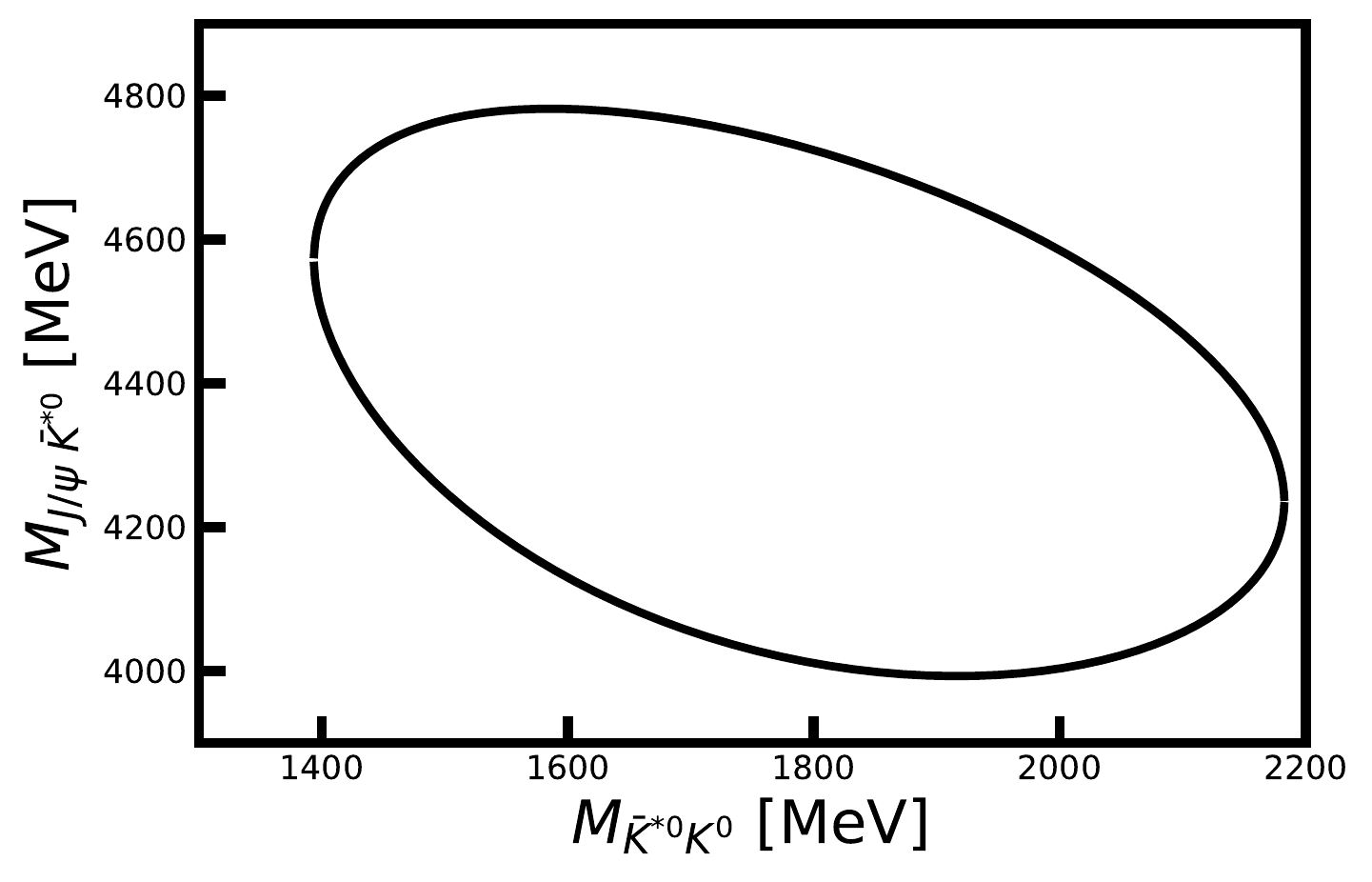}
\caption{Dalitz plot for $\bar{K}^{*0} K^0$ and $J/\psi \bar{K}^{*0}$ invariant masses in the $\bar{B}^0 \to J/\psi \bar{K}^{*0} K^0$ decay.}
\label{Dalitz}
\end{figure}

In the Eq.(\ref{eq20}), $M_{\bar{K}^{*0} K^0}$ and $M_{J/\psi, \bar{K}^{*0}}$ are interrelated, 
thus for specific $M_{\bar{K}^{*0} K^0}$, the range of $M_{J/\psi, \bar{K}^{*0}}$ is defined as 
\begin{small}
\begin{eqnarray}\label{eq22} \nonumber
  && M_{J/\psi, \bar{K}^{*0}}^{max}= \\ \nonumber
  && \sqrt{(E_{\bar{K}^{*0}}+E_{J/\psi})^2-\left(\sqrt{E_{\bar{K}^{*0}}^2-M_{\bar{K}^{*0}}^2}-\sqrt{E_{J/\psi}^2-M_{J/\psi}^2} \right)^2},\\ \nonumber
  && M_{J/\psi, \bar{K}^{*0}}^{min}= \\ \nonumber
  && \sqrt{(E_{\bar{K}^{*0}}+E_{J/\psi})^2-\left(\sqrt{E_{\bar{K}^{*0}}^2-M_{\bar{K}^{*0}}^2}+\sqrt{E_{J/\psi}^2-M_{J/\psi}^2} \right)^2}, \\
\end{eqnarray}
\end{small}
where $E_{\bar{K}^{*0}}=(M_{inv}^2-M_{K}^2+M_{\bar{K}^{*0}}^2)/2M_{inv} $ and $ E_{J/\psi }=(M_{\bar{B}^0}^2-M_{inv}^2-M_{J/\psi}^2)/2M_{inv}$ 
are the energies of $\bar{K}^{*0}$ and $J/\psi$ in the $\bar{K}^{*0} K^0$ rest frame. 
The Dalitz plot for $M_{\bar{K}^{*0} K^0}$ and $M_{J/\psi \bar{K}^{*0}}$ invariant masses 
in the $\bar{B}^0 \to J/\psi \bar{K}^{*0} K^0$  decay is shown in FIG.\ref{Dalitz}.

On the other hand, notice that Eq.(\ref{eq20}) contains an unknown quantity $V_P$. 
In order to eliminate $V_P$, we could consider the production of the $f_1(1285)$ resonance. 
The relevant mechanism is depicted in FIG.\ref{fig2}(c) and we have
\begin{equation}\label{eq23}
\begin{aligned}
T_{\bar{B}^0 \to J/\psi f_1(1285)} = V_P C_s' G(M_{f_1}) g_{f_1}
\end{aligned}
\end{equation}
where the polarization factor $C_s'$ is obtained analogy to Eq.(\ref{eq21}). 
The coupling of $f_1(1285)$ to $K^* \bar{K}$ channel can be expressed as $g_{f_1} \epsilon_i(f_1) \epsilon_i(K^*)$,
by contracting the two polarization vectors $\epsilon_i(K^*)\epsilon_j(K^*)=\delta_{ij}$ in the loop of FIG.\ref{fig2}.(c), and give
\begin{eqnarray}\label{eq24}
\sum C_s'^2= 3+\frac{\vec{p}_{J/\psi}^2}{m_{J/\psi}^2}+\frac{\vec{p}_{f_1}^2}{m_{f_1}^2}.
\end{eqnarray}
where $\vec{p}_{f_1}=\vec{p}_{J/\psi}=\lambda^{\frac{1}{2}}(M_{\bar{B}^0}^2,M_{J/\psi}^2,M_{f_1}^2)/2M_{\bar{B}^0}$ is the three momentum of $f_1$ and $J/\psi$ in the $\bar{B}^0$ rest frame.

Then, the partial decay width of $\bar{B}^0 \to J/\psi f_1(1285)$ can be written
\begin{eqnarray}\label{eq25} \nonumber
\Gamma_{\bar{B}^0 \to J/\psi f_1}&=&\frac{V_P^2}{8\pi}\frac{g_{f_1}^2 G^2(M^2_{f_1})  |\vec{p}_{J/\psi}|}{M_{\bar{B}^0}^2}\\
 &\times& \left(2+\frac{\left( M_{\bar{B}^0}^2-M_{J/\psi}^2-M_{f_1}^2 \right)^2}{4M_{J/\psi}^2M_{f_1}^2} \right)
\end{eqnarray}

By considering $\bar{B}^0 \to J/\psi f_1(1285)$, we define
\begin{eqnarray}\label{eq26}
R_\Gamma =\frac{d \Gamma_{\bar{B}^0 \to J/\psi \bar{K}^{*0} K^0}/dM_{inv}(\bar{K}^{*0} K^0)}{\Gamma_{\bar{B}^0 \to J/\psi f_1(1285)}}
\end{eqnarray} 
where $V_P$ has been cancelled so that $R_\Gamma$ has no other free parameters except the energy scale. 
It can be a function of invariant mass $M_{inv}$ and compared with the experiments at a certain energy scale. 
Although there are different $\mathcal{U}$ for different final states $J/\psi \bar{K}^{*0} K^0$ and $J/\psi K^{*0}\bar{ K}^0$, 
they have an equal $R_\Gamma$.

\section{NUMERICAL RESULTS AND DISCUSSION}
\label{IV}

\begin{figure*}[]
\centering
\subfigure[]{
\includegraphics[width=0.3\linewidth]{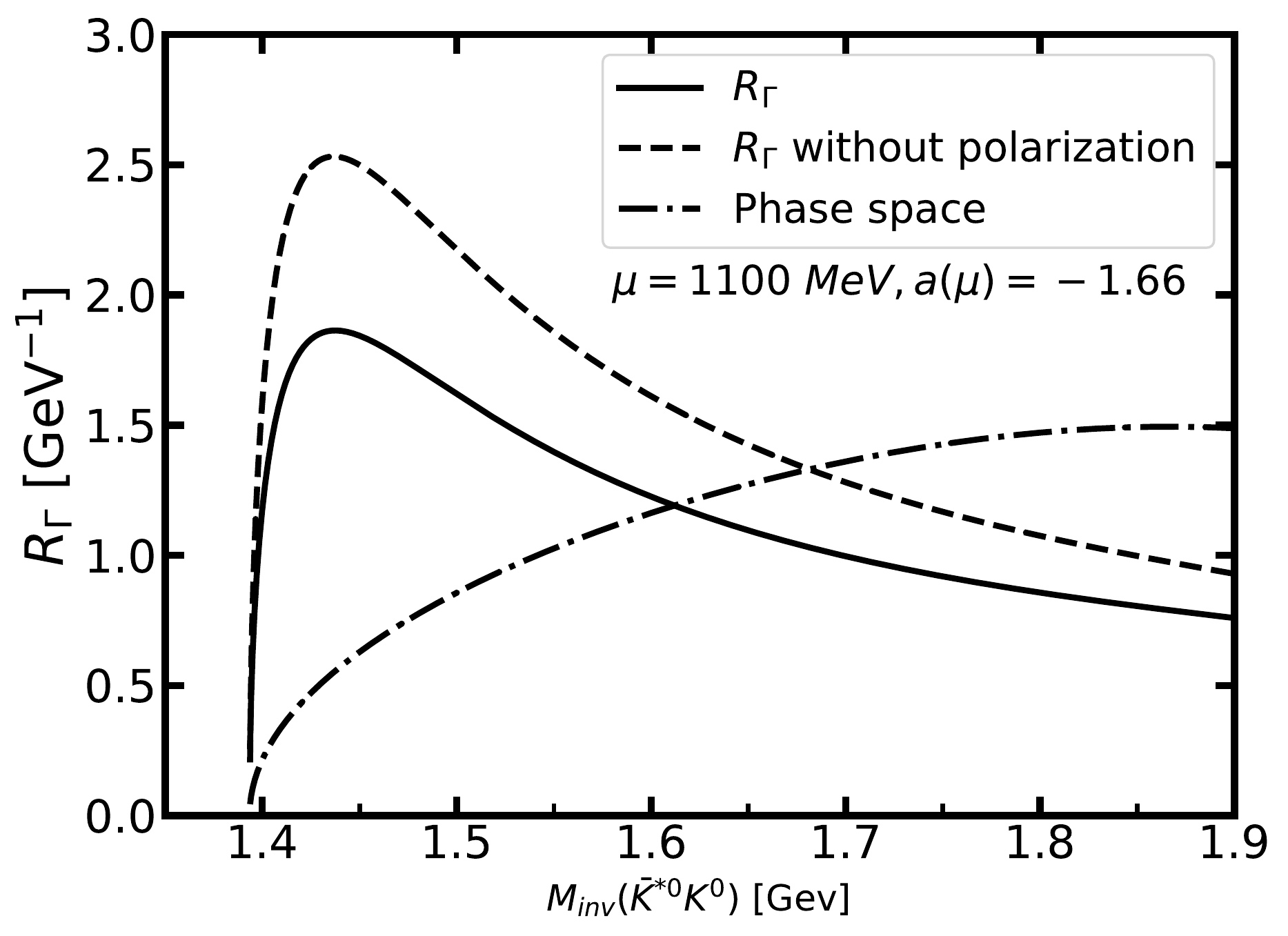}}
\hspace{0.01\linewidth}
\subfigure[]{
\includegraphics[width=0.3\linewidth]{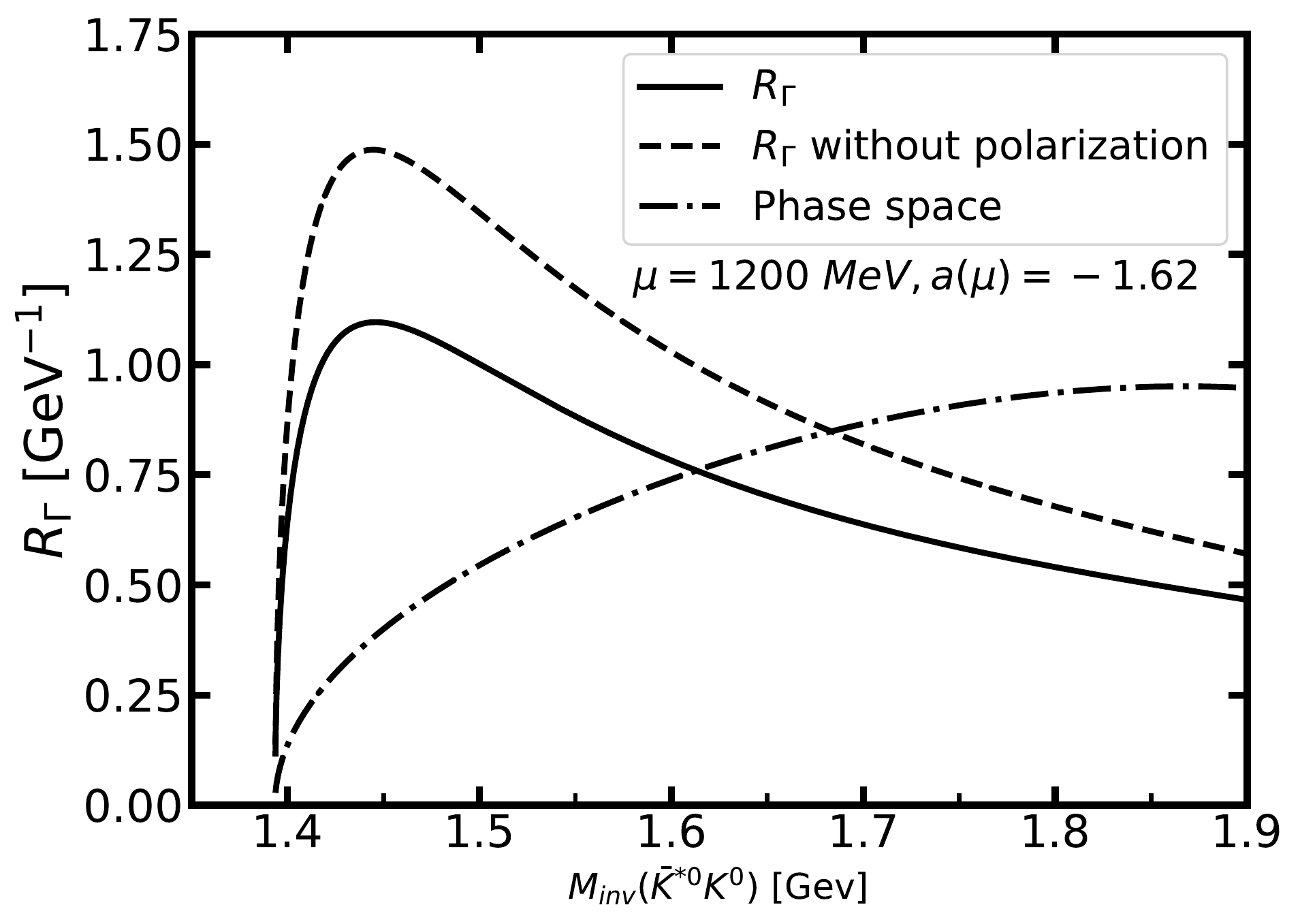}}
\subfigure[]{
\includegraphics[width=0.3\linewidth]{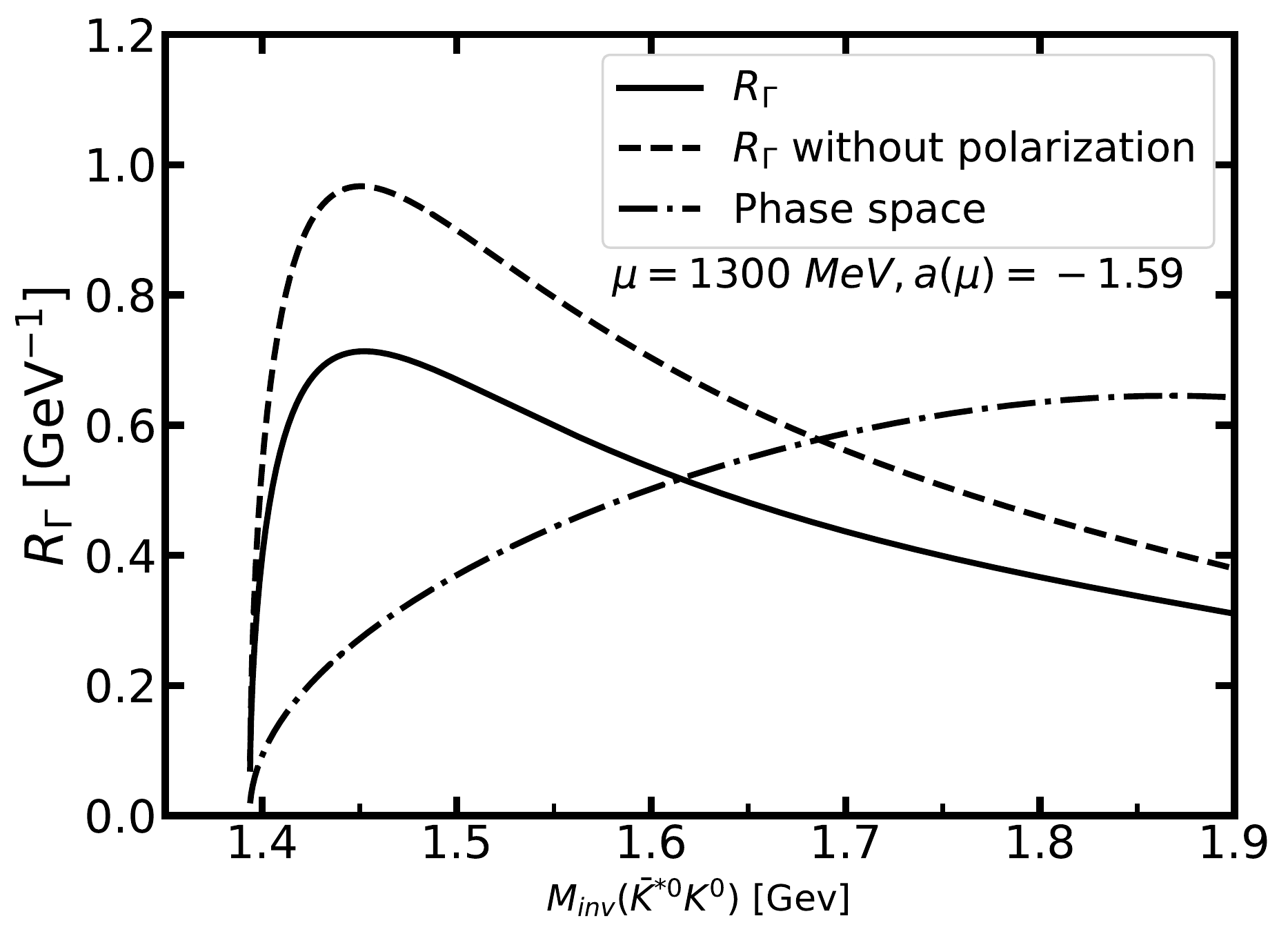}}
\hspace{0.01\linewidth}
\caption{
The distribution of $R_{\Gamma}$ as a function of the invariant mass $M_{inv}$ of $\bar{K}^{*0} K^0$ at  $\mu = 1100\text{MeV},\ 1200\text{MeV}$ and $ 1300\text{MeV}$, respectively.
The solid (dashed) curve represents the dynamically generated prediction with (without) the polarization structure factor.
The dot-dashed curve stands for the phase space distribution which has been normalized  to make it have the same area with the solid curve in the $M_{inv}$ range.}
\label{fig4}
\end{figure*}

With the former  formulae, we present, in FIG.\ref{fig4}, the predictions for the ratio $R_\Gamma$ as a function of the invariant mass $M_{inv}$ of the $\bar{K}^{*0} K^0$ state.  
As mentioned above, the unknown parameter $V_P$ cancels, therefore it does not appear in the final result of the ratio $R_\Gamma$. 
Meanwhile, we consider the following renormalization scales $\mu = 1100\text{MeV},\ 1200\text{MeV}, \ 1300\text{MeV}$  
and corresponding subtraction constants $a(\mu) = -1.66,\ -1.62, \ -1.59$. 
A straightforward relation between $a(\mu)$ and $\mu$ is given by \cite{Oller:2000fj}
\begin{equation}
a_i(\mu)=-2\log{\left(1+\sqrt{1+\frac{M_i^2}{\mu^2}} \right)}+\mathcal{O}\left(\frac{m_i}{M_i} \right).
\end{equation}

As shown in FIG.\ref{fig4}, the solid curve gives a good description for the invariant mass distribution of $M_{inv}(\bar{K}^{*0} K^0)$, 
which produces the broad peak of the mass distribution. 
The broad peak is caused by the $f_1(1285)$ which is dynamically generated by the $K^* \bar{K}- c.c.$ interaction. 
The range of the invariant mass $M_{inv}$ of $\bar{K}^{*0} K^0$ has been considered as the $1390-1900$ MeV. 
In principle, the range of the invariant mass $M_{inv}$ of $K^{*}\bar{K}$ can reach the maximum value $2182$ MeV($M_{\bar{B}^0}-M_{J/\psi})$ 
in the process of $\bar{B}^0 \to J/\psi \bar{K}^{*0} K^0$ decay.  Nevertheless, 
the results of the previous sections are based on the chiral unitary theory, 
thus the range of the invariant mass should be $200-300$ MeV smaller than the maximum value $2182$ MeV to make sure the theory works better \cite{Xie:2015lta}.
At the same time, by comparing three subgraphs in FIG.\ref{fig4}, we can see that as the renormalization scale $\mu$ increases, 
the strength of the  $R_\Gamma$ decreases and the position of the broad peak of $R_\Gamma$ almost does not move.
In addition, in order to compare, the result without considering the polarization structure factor of the $R_{\Gamma}$ is also shown by the dashed curve.
We can see that the dashed curve also gives a clear broad peak and is larger than the solid one.
Finally, the phase space as a function of the invariant mass $M_{inv}$ has been depicted by the dot-dashed curve. 
Actually, the value from the phase space distribution is much smaller than the values of the other two curves, 
and we normalize the phase space distribution to make it have the same area with the solid curve.  

We define the parameter $R_t$ by 
\begin{eqnarray}\label{Rt}
R_t = \frac{\Gamma_{\bar{B}^0 \to J/\psi \bar{K}^{*0} K^0}}{\Gamma_{\bar{B}^0 \to J/\psi f_1(1285)}} = \int_{m_{\bar{K}^{*0}}+m_{K^0}}^{m_{\bar{B}^0}-m_{J/\psi}} R_\Gamma \ d M_{inv},
\end{eqnarray} 
and obtain a theoretical value of $R_t=0.451$ for $\mu=1200$ MeV. 
When we choose the values of  $\mu$ in the range of $1100 \sim 1300$ MeV, 
the results of $R_\Gamma$ are located around the range of  $0.301 \sim  0.734$. 
Thus the main uncertainty of the theoretical ratio comes from the renormalization scale $\mu$.

On the other side, the processes $K^{*0} \to K^+ \pi^-,K^0\pi^0$ have strengths $\frac{2}{3}$, $\frac{1}{3}$, respectively. 
The strength of the process $\bar{B}^0 \to J/\psi K^*\bar{K}$ can be obtained by using the experimental branching ratio of $\bar{B}^0 \to J/\psi K^0 K^- \pi^+ +c.c$, 
thus similar to Ref.\cite{Liang:2017ijf}, we have
\begin{eqnarray}\label{eq27} \nonumber
&&Br(\bar{B}^0 \to J/\psi \bar{K}^{*0} K^0)=\frac{1}{8}  \frac{3}{2}\times 2.1 \times 10^{-5},\\
&&Br(\bar{B}^0 \to J/\psi f_1(1285))= (8.4 \pm 2.1 ) \times 10^{-6},
\end{eqnarray}
where the branching ratio for $\bar{B}^0 \to J/\psi f_1(1285)$ can be obtained directly from PDG. The Eq.(\ref{eq27}) gives an experiment result
\begin{equation}\label{Re}
R_e =\frac{Br(\bar{B}^0 \to J/\psi \bar{K}^{*0} K^0)}{Br(\bar{B}^0 \to J/\psi f_1(1285))}= 0.469 \pm 0.117.
\end{equation}
In principle, the $R_e$ should be related with $R_t$ by $R_e=R_t$.
However, the $Br(\bar{B}^0 \to J/\psi K^0 K^- \pi^+ +c.c.) = 2.1 \times 10^{-5}$ is not accurate within the corresponding experimental condition. 
Furthermore, as mentioned, in order that the chiral unitary approach can work well, 
the upper limits of the $M_{inv}$ should be $200 - 300$ MeV far from the boundary($m_{\bar{B}^0}-m_{J/\psi}$). 
The above two reasons lead to the main discrepancy between $R_e$ and $R_t$. 
There are also some other reasons, for examples, there are higher mass states with spin-parity $J^P=1^+$ and $2^+$ at higher invariant mass range of $\bar{K}^{*0}K^0$ \citep{Xie:2015lta}. 
As for the relation between $\mu$ and $a(\mu)$, we ignore higher-order effects. 
These contributions are small compared with the previous two main reasons.
We expect that future experimental observation of the invariant mass distribution $R_\Gamma$ on the LHCb  
would provide valuable information on the mechanism of the $\bar{B}^0 \to J/\psi \bar{K}^{*0} K^0$ decay.

\section{SUMMARY}
\label{V}
In the present work, we have shown the distribution of the invariant mass $\bar{K}^{*0} K^0$ in the decay $\bar{B}^0 \to J/\psi \bar{K}^{*0} K^0$ 
to investigate the basic nature of the $f_1(1285)$ resonance. In the frame of chiral unitary theory, the $f_1 (1285)$ resonance is dynamically generated 
from the $K^*\bar{K}-c.c.$ interaction, further we get the theoretical prediction 
$R_\Gamma = \frac{d \Gamma_{\bar{B}^0 \to J/\psi \bar{K}^{*0} K^0}/dM_{inv}(\bar{K}^{*0} K^0)}{\Gamma_{\bar{B}^0 \to J/\psi f_1(1285)}}$ 
as a function of invariant mass $M_{inv}(\bar{K}^{*0} K^0)$. The result of $R_\Gamma$ reproduces the peak of the mass distribution of $M_{inv}(\bar{K}^{*0} K^0)$, 
which is caused by the production of the $f_1 (1285)$ state. 
Finally, we can obtain a theoretical result $R_t=\Gamma_{\bar{B}^0 \to J/\psi \bar{K}^{*0} K^0}/\Gamma_{\bar{B}^0 \to J/\psi f_1(1285)}$ 
which is possible to be compared with the experimental result $R_e$. 
We expect that there will be more valuable data in future experimental observations, which could advance our understanding of $f_1(1285)$.

\begin{acknowledgments}
The authors thank Eulogio Oset for his valuable comments. 
Hao Sun is supported by the National Natural Science Foundation of China (Grant No.12075043).
\end{acknowledgments}

\bibliography{ref}
\end{document}